\documentclass[aip,pre,reprint,superscriptaddress,twocolumn]{revtex4-1}
\usepackage{graphicx}% Include figure files
\usepackage[english]{babel} % 
\usepackage[usenames]{color}
\usepackage{afterpage} 
\usepackage{subfigure}
\usepackage{dcolumn}% Align table columns on decimal point
\usepackage{bm}% bold math
\usepackage{amsmath,amsfonts,amssymb,float}
\usepackage{natbib}
\usepackage{epsfig}
\usepackage{textcomp} % allows upright mu s%
\usepackage{amsmath} % AMS Math Package
\usepackage{amsthm} % Theorem Formatting
\usepackage{amssymb}	% Math symbols such as \mathbb
\usepackage{graphicx} % Allows for eps images
\makeatletter % Need for anything that contains an @ command 
\let\baraccent=\= % rename builtin command \= to \baraccent
\renewcommand{\=}[1]{\stackrel{#1}{=}} % for putting numbers above =

\theoremstyle{definition}

\theoremstyle{remark}

\begin{document}

%\preprint{}

\title{Measuring fast electron spectra and laser absorption in relativistic laser-solid interactions using differential bremsstrahlung photon detectors}

\author{R.H.H. Scott}
\email{Robbie.Scott@stfc.ac.uk}
\thanks{The authors gratefully thank the staff of the LULI2000 laser facility, Ecole Polytechnique, Paris, France. The author would also like to thank Stuart Ansel for useful \sc mcnpx \rm discussions. This investigation was undertaken as part of the HiPER preparatory project and was funded by LaserLab Europe, Ecole Polytechnique and by the Science and Technology Facilities Council, UK.}

\affiliation{Central Laser Facility, STFC, Rutherford Appleton Laboratory, Harwell Oxford, Didcot, OX11 0QX, United Kingdom}
\affiliation{Department of Physics, Blackett Laboratory, Imperial College London, Prince Consort Road, London, SW7 2AZ, United Kingdom}
\author{E.L. Clark}
\affiliation{TEI of Crete,Romanou 3,Chania 73133,Greece}
\author{F. P\'erez}
\affiliation{LULI, \'Ecole Polytechnique, UMR 7605, CNRS/CEA/UPMC, Route de Saclay, 91128 Palaiseau, France}
\author{M.J.V. Streeter}
\affiliation{Department of Physics, Blackett Laboratory, Imperial College London, Prince Consort Road, London, SW7 2AZ, United Kingdom}
\author{J.R. Davies}
\affiliation{GoLP, Instituto de Plasmas e Fus\~ao Nuclear - Laborat\'orio Associado, Instituto Superior T\'ecnico, 1049-001 Lisboa, Portugal}
\affiliation{Fusion Science Center for Extreme States of Matter, University of Rochester, Rochester, New York 14623, USA}
\affiliation{Laboratory for Laser Energetics, University of Rochester, Rochester, New York 14623, USA}
\affiliation{Department of Mechanical Engineering, University of Rochester, Rochester, New York 14623, USA}
\author{H.-P. Schlenvoigt}
\affiliation{LULI, \'Ecole Polytechnique, UMR 7605, CNRS/CEA/UPMC, Route de Saclay, 91128 Palaiseau, France}
\author{J.J. Santos}
\affiliation{Univ. Bordeaux/CNRS/CEA, CELIA, UMR 5107, 33405 Talence, France}
\author{S. Hulin}
\affiliation{Univ. Bordeaux/CNRS/CEA, CELIA, UMR 5107, 33405 Talence, France}
\author{K.L. Lancaster}
\affiliation{Central Laser Facility, STFC, Rutherford Appleton Laboratory, Harwell Oxford, Didcot, OX11 0QX, United Kingdom}
\author{S.D. Baton}
\affiliation{LULI, \'Ecole Polytechnique, UMR 7605, CNRS/CEA/UPMC, Route de Saclay, 91128 Palaiseau, France}
\author{S.J. Rose}
\affiliation{Department of Physics, Blackett Laboratory, Imperial College London, Prince Consort Road, London, SW7 2AZ, United Kingdom}
\author{P.A. Norreys}
\affiliation{Central Laser Facility, STFC, Rutherford Appleton Laboratory, Harwell Oxford, Didcot, OX11 0QX, United Kingdom}
\affiliation{Department of Physics, Blackett Laboratory, Imperial College London, Prince Consort Road, London, SW7 2AZ, United Kingdom}

\date{\today}

\begin{abstract}
A photon detector suitable for the measurement of bremsstrahlung spectra generated in relativistically-intense laser-solid interactions is described. The Monte Carlo techniques used to back-out the fast electron spectrum and laser energy absorbed into forward-going fast electrons are detailed. A relativistically-intense laser-solid experiment using frequency doubled laser light is used to demonstrate the effective operation of the detector. The experimental data was interpreted using the 3-spatial-dimension Monte Carlo code \sc mcnpx\rm (Pelowitz 2008), and the fast electron temperature found to be 125 keV. 

\end{abstract}

%\keywords{}

\maketitle
\section{Introduction}

When an intense laser pulse irradiates a solid target, large numbers of fast electrons are generated. The kinetic energy these fast electrons acquire during the interaction has been found to be close to that of electrons oscillating in the transverse field of the incident light wave \cite{Wilks:1992nq, Sherlock:2009aa, haines:045008}.  If the picosecond-duration laser pulse is associated with a much longer duration, lower intensity pedestal (caused by amplified spontaneous emission), then a coronal plasma is formed around the interaction point some time before the relativistic pulse arrives. In these circumstances, it is possible for the fast electrons to acquire substantially higher energies \cite{key:1966,scott:053104} via collective acceleration processes in the finite density gradient plasma. 

The understanding of fast electron energy transport caused by intense laser-plasma interactions is an interesting area of fundamental physics. Furthermore the detailed characterization of the fast electron transport under controlled conditions is important for numerous applications including proton and ion beam production \cite{refId,PhysRevLett.84.670, PhysRevLett.85.2945, PhysRevLett.84.4108}, isochoric heating of high density matter for opacity studies \cite{Hoarty2007115, PhysRevLett.91.125004, PhysRevLett.104.085001, nishimura:022702, PhysRevLett.82.4843}, and fast ignition inertial confinement fusion \cite{Tabak:1994ov,Kodama:2001ss,Honrubia:2006mb,0741-3335-51-1-015016,Evans:2007xr,PhysRevLett.77.2483}. Fast electron energy transport is dictated by the resistivity and density of the medium in which the transport is occurring, the fraction of laser energy absorbed into fast electrons, the fast electron source size, their divergence and energy spectrum. 

Measurements of the fast electron energy spectrum can be made using magnetic spectrometers\cite{PhysRevLett.77.75,tanaka:2014}, nuclear activation\cite{stoyer:767}, bremsstrahlung x-ray measurements\cite{Beg:1997pd,stoyer:767,chen:10E305,courtois:013105,hatchett:2076,key:1966}, buried fluorescent foils\cite{yasuike:1236} and proton emission \cite{Beg:1997pd}. However interpretation of the measurements is can be complicated by the formation of a large electro-static field at the target rear surface/vacuum boundary by the initial population of electrons exiting the target \cite{wilks:542}. This field confines the subsequent fast electrons, meaning only a small fraction of the fast electrons accelerated by the laser are able to exit the target rear. The spectrum of the fast electrons which do exit the target can be measured relatively simply using magnetic spectrometers, however as the bulk of the electrons remain within the target, it is likely that the fast electron spectrum measured from this initial electron population is not representative of the bulk fast electron spectrum. Bremsstrahlung photon emission, caused by fast electron collisions with the solid target, offers a mechanism by which the spectrum and total laser absorption of the bulk fast electron population can be inferred. 

This paper describes the design of bremsstrahlung detectors and the interpretation of experimental data using Monte Carlo techniques. Preliminary analysis of the fast electron transport induced by frequency doubled, relativistically-intense laser-solid interactions is presented. Section \ref{sec:canondesignprinciples} begins by discussing the bremsstrahlung generation mechanism and previously employed bremsstrahlung measurement techniques, while section \ref{sec:canondesign} goes on to describe the design of a novel bremsstrahlung detector. Section \ref{sec:bremmodel3} describes how Monte Carlo modelling can be used to back-out the fast electron temperature, divergence and total laser energy absorbed into fast electrons. In order to demonstrate the functionality of this new detector, section \ref{expt} briefly describes a frequency doubled, relativistically-intense laser solid-interaction, from which the fast electron temperature is inferred using Monte Carlo techniques.

\section{Background}
\label{sec:canondesignprinciples}
The energy of an emitted bremsstrahlung photon is dictated by the energy the fast electron loses during the deceleration associated with a collision, and is therefore related to both the angle through which the electron is scattered and its initial energy. The photon's emission direction is limited to a cone, the major axis of which is the initial propagation direction of the incident electron, while the opening angle of the cone is determined by the initial Lorentz factor of the fast electron. The relationship between the incident electron and bremsstrahlung photon mean that absolutely calibrated measurements of the spatial and spectral distribution of the bremsstrahlung emission can enable the determination of the fast electron spectrum, laser absorption fraction and fast electron spatial distribution. Due to the number of particles involved, interpretation of experimental data rapidly becomes extremely complex meaning data analysis is greatly simplified by the use of Monte Carlo type simulations. 

\begin{figure}[ht]
\includegraphics[scale=0.38,trim=10mm 65mm 25mm 80mm,clip]{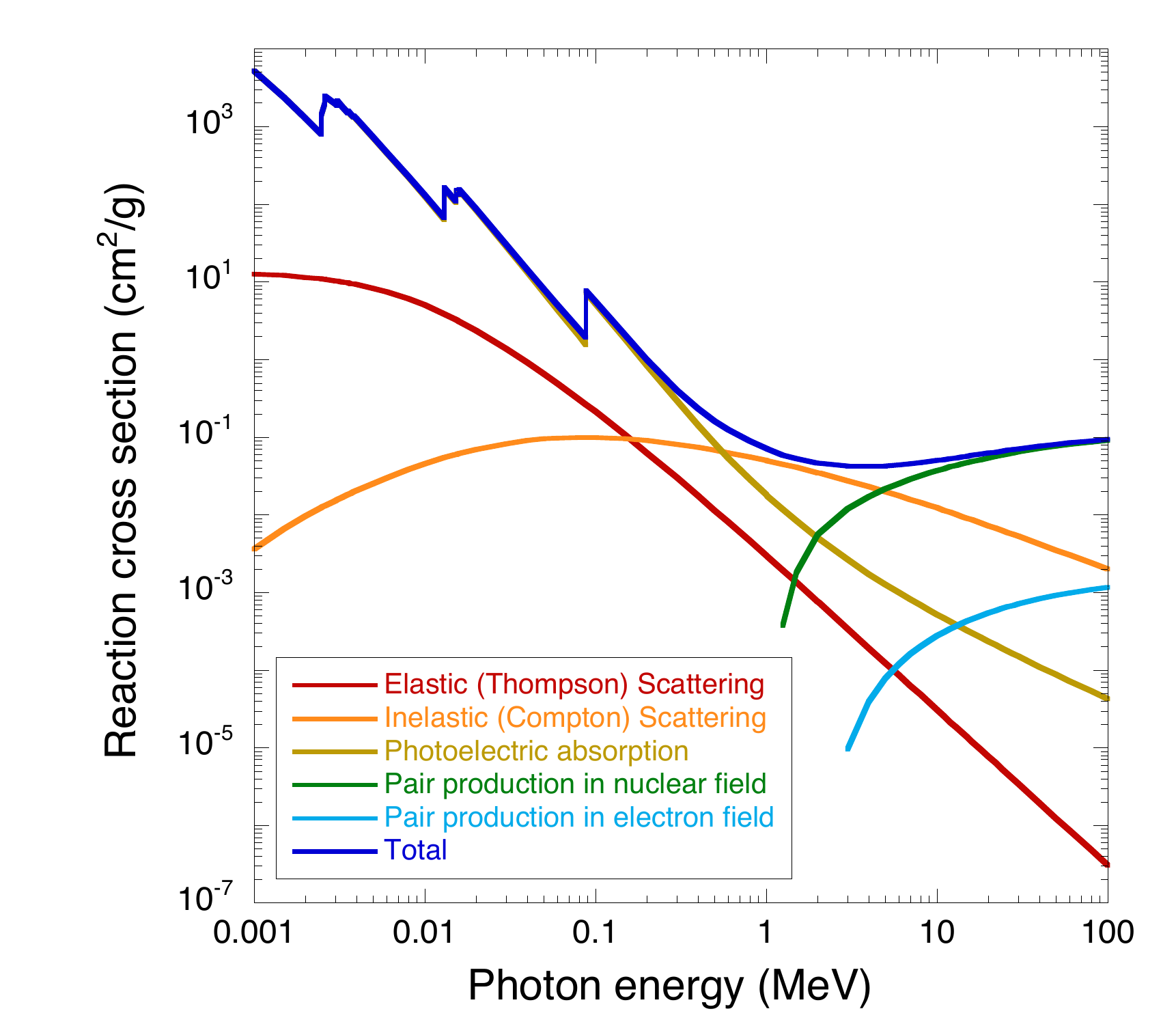}
\caption{Reaction cross-sections with respect to incident photon energy for Pb, source XCOM, NIST. Photonuclear interactions are not shown but are extremely small for Pb for photon energies $< 10$ MeV.}
\label{fig:fepbcrosssections}
\end{figure}

Bremsstrahlung radiation measurement techniques typically employ either differential filtering or photonuclear activation. Photonuclear techniques \cite{stoyer:767,hatchett:2076} are limited to photons of energies $> 5$ MeV due to the rapid decrease in the photonuclear cross-section below this energy. For the purposes of fast ignition inertial confinement fusion, the electron energies of interest are those between $1-5$ MeV hence photonuclear techniques are less well suited to such measurements. Differential filtering \cite{Beg:1997pd,chen:10E305,courtois:013105,key:1966} works by using many filters of differing thicknesses and measuring the flux exiting the filters. This technique relies on spectral variations in opacity; away from resonance regions such as the K edges of materials, the opacity decreases monotonically with incident photon energy, typically followed by a slight increase above $\sim1-5$ MeV (figure \ref{fig:fepbcrosssections}) due to the increasing photonuclear cross-section. As photons propagate through matter, the flux reduces exponentially with increasing distance, while from figure \ref{fig:fepbcrosssections} the cross-section is generally inversely proportional to photon energy. Therefore as the photons propagate through a given material, the low energy flux undergoes more collisions (and ultimately absorption) than the high energy photon flux. The change in cross-section with material can also potentially be used advantageously, for example in a Ross pair configuration the K-edges of various materials are used to preferentially remove photons with energies closely corresponding to that of the K-edge of the filter material. However this method is only effective for photon energies up to $\sim$100 keV, as although the K-edge energy increases with atomic number, even Uranium only has a K-edge of 115 keV. Photon flux is generally measured indirectly via photon energy deposition, various different techniques exist such as thermoluminescent detectors, image plates and pin diodes. 

Accurate interpretation of differential filtering is complicated as photon collisions lead to the emission of secondary particles through various mechanisms such as the photoelectric effect, pair production, or photonuclear interactions. These secondary particles may in turn emit their own secondaries, hence the situation rapidly becomes very complex. Monte Carlo modelling is well suited to this type of problem, allowing a quantitative measure of the signal to be inferred. 

Bremsstrahlung detectors have been used in numerous guises on laser-solid experiments in the past \cite{Beg:1997pd,stoyer:767,chen:10E305,courtois:013105,hatchett:2076,key:1966}. Limitations of previous detectors have generally arisen from the relatively small number of energy bins and the inability to measure higher energy photons.  

Beg \emph{et al} \cite{Beg:1997pd} used bremsstrahlung measurements to diagnose the hot electron temperature. These measurements were made using an array of pin diodes with photomultiplier/scintillator detectors in combination with substantial Pb collimation, shielding and 2-5 mm of differential Pb filtering. This setup enabled the measurement of the relationship $T_{hot}\sim 100(I\lambda^2)^{1/3}$, where $T_{hot}$ is the slope temperature of the exponential energy spectrum in keV, $I$ the laser intensity in units of $1\times10^{17}$ W/cm$^2$ and $\lambda$ the laser wavelength in \textmu m. 

Li$_{2}$B$_{4}$O$_{7}$ and CaSO$_{4}$ thermoluminescent detectors were used by Key \emph{et al} to detect bremsstrahlung emission \cite{key:1966}.  

More recently Chen \emph{et al} \cite{chen:10E305} designed a bremsstrahlung detector primarily based on differential filtering using 13 different materials which separated image plates. The materials were chosen so that their K-edges lay in an appropriate part of the spectrum thereby preferentially filtering photons of that energy. In combination with \sc Integrated Tiger Series \rm Monte Carlo modelling of the electron and photon transport, this techniques allows good resolution of the lower energy region of the spectrum. This method provides good spectral resolution in the spectral region examined, but has the practical disadvantage of the number of image plates which are required; extracting, scanning and replacing 13 image plates after each laser shot creates a huge experimental workload limiting the number of detectors which can be fielded on an experiment and hence the spatial resolution. 

\section{Design of a novel bremsstrahlung detector using \sc mcnpx \rm}
\label{sec:canondesign}
A new bremsstrahlung detector design was conceived with the aim of providing good energy resolution across the spectral region of interest, whilst allowing numerous detectors to be fielded on one experiment. The ability to field multiple detectors is important as it allows the angular emission of the bremsstrahlung emission to be measured. As the cone angle of the bremsstrahlung emission is related to the fast electron energy, the angular resolution provides a secondary measurement with which to further constrain Monte Carlo calculations. From these measurements it is theoretically possible to infer the fast electron temperature as a function of angle, the absolute number of fast electrons and hence the fraction of laser energy absorbed into fast electrons, and potentially the fast electron divergence angle. 

\begin{figure}[ht]
\centering
\includegraphics[scale=0.35,angle=0,trim=170mm 65mm 2mm 25mm,clip]{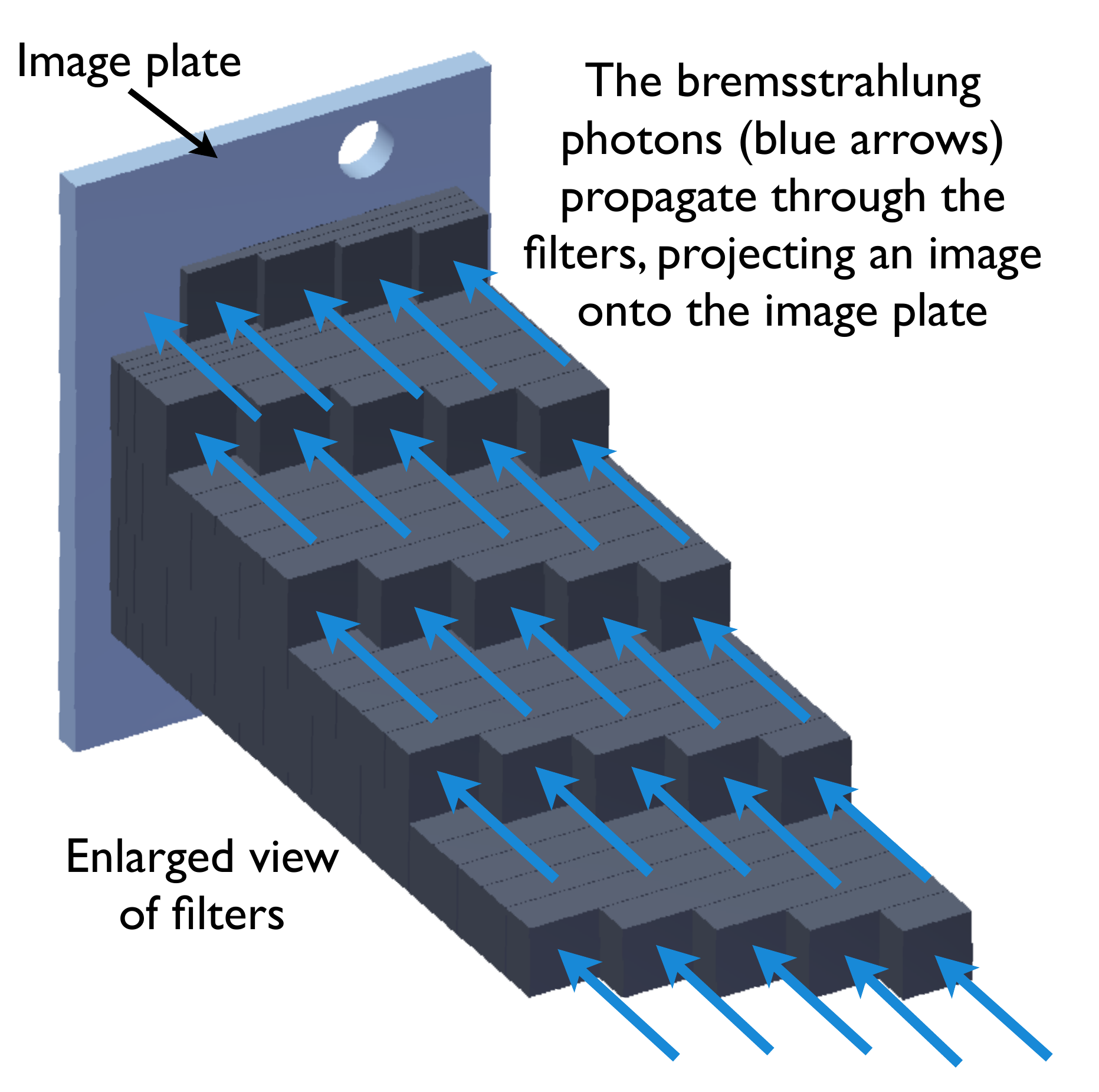}
\caption{The incident bremsstrahlung photons (blue arrows) propagate through the filter array, creating an image on the region of image plate behind the filters, this image contains convolved information about the spectral distribution of the bremsstrahlung photons. The 25 filters create 25 energy bins.}
\label{brem3D}
\end{figure}

The filtering was chosen to be of differential design in order that the majority of $< 3$ MeV photons emitted by the electrons of importance to fast ignition can be resolved. The filters were comprised of just one material with differing filter thicknesses. A design solution was arrived at which only requires one image plate - a significant improvement over previous designs. The filter design is shown in figure \ref{brem3D}, 25 different thicknesses of Pb filtering in a parallel configuration filter the incident x-ray spectrum. The transverse dimensions of each filter are 0.5$\times$0.5 cm. The transmitted photons deposit their energy within the image plate positioned behind the filters, the energy deposited is a function of the bremsstrahlung spectrum, filter thickness, filter material and image plate response. This filtering concept requires that the bremsstrahlung flux is effectively spatially and spectrally uniform across the cavity of the detector. Modelling showed that this assumption is valid as long as the detector is sufficiently far from the electron source, 40 cm is more than sufficient. 

\begin{figure*}[ht]
\centering
\includegraphics[scale=1,trim=55mm 20mm 40mm 27mm,clip]{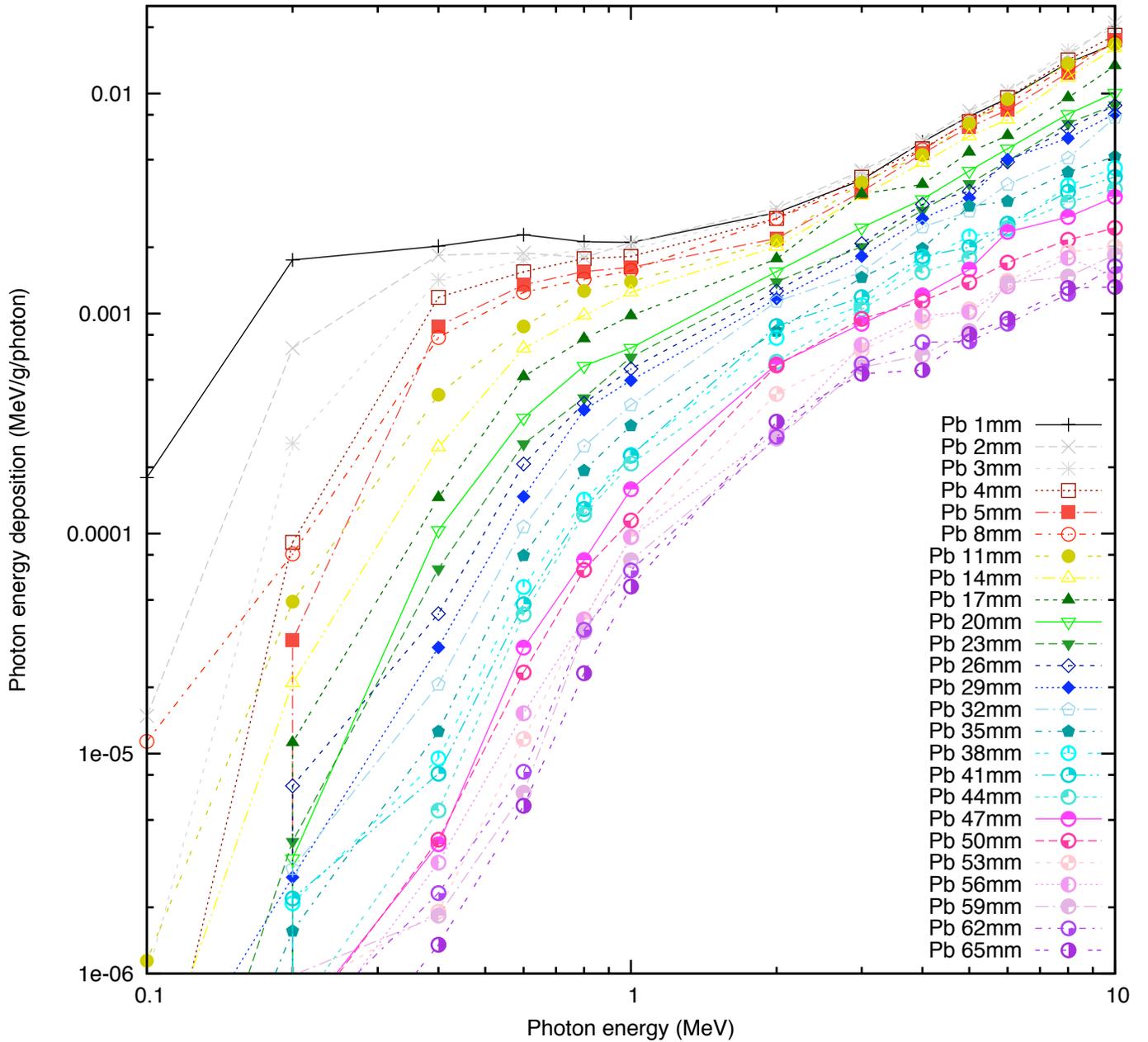}
\caption{Energy deposition within the active phosphor layer of the image plate as a function of photon energy and Pb filter thickness, as modelled with \sc mcnpx\rm. Each line represents a different thickness of Pb filtering, the thicknesses shown are those which were eventually used on the final design. The relative errors are not shown for clarity, but it can be seen from the erratic nature of the curves that the relative error is higher when the photon energies are lowest and the filters thicker. The response curves indicate the detector will be able to distinguish between photons of energy $\leq 2-3$ MeV.}
\label{fig:pbfiltering}
\end{figure*}

Detailed design of the bremsstrahlung detector was performed by using \sc mcnpx\rm \cite{Pelowitz:aa} to solve the 3 dimensional (3D) coupled electron-photon transport problem. Excluding the effects of self-generated electro-magnetic fields, and using the assumption that the cold cross-sections remain accurate at temperatures of $\sim 50$ eV, \sc mcnpx \rm can accurately model the electron transport and bremsstrahlung generation via collisions within the solid target. The subsequent photon flux propagation through the filters and energy deposition within the image plate can also be modelled as one self-consistent problem, whilst accounting for undesirable/complex effects such as K$_{\alpha}$ emission and pair production within the filters. A detailed characterization of Pb and Fe as filtering materials was performed as they both have relatively high total cross-sections, also the minimum inflection point of their total cross-sections is $> 1$ MeV which should allow for spectral resolution up to this point. 3D \sc mcnpx \rm modelling was used to quantify the photon energy deposition within the active phosphor layer of the image plate as a function of both photon energy and filter thickness. A 3D filtering array was created spanning a wide filtering thickness range within \sc mcnpx\rm. In order to characterize photon energy deposition as a function of photon energy, numerous mono-energetic photon source runs were performed, yielding the energy deposition as a function of filter thickness for that particular photon energy. The data from many runs was aggregated to create figure \ref{fig:pbfiltering}. The response curves indicate the detector will be able to distinguish between photons of energy $\leq 2.5$ MeV (note this is the energy of individual photons, not the fast electron slope temperature), above 2.5 MeV the response of the various channels is indistinguishable (i.e. they have the same slope) due to the minimum in the Pb cross-section (figure \ref{fig:fepbcrosssections}). The thicknesses shown are those used on the final design.

Teflon and other low $Z$ materials were excluded as structural materials as they were found to greatly reduce the sensitivity of the detector by flooding the image plate with low energy K$_{\alpha}$ photons, creating a high background signal. Modelling showed high $Z$ materials were better in this respect, hence Pb was used structurally throughout.  

\begin{figure}[ht!]
\centering
\includegraphics[scale=0.44,trim=0.5mm 08mm 130mm 10mm,clip]{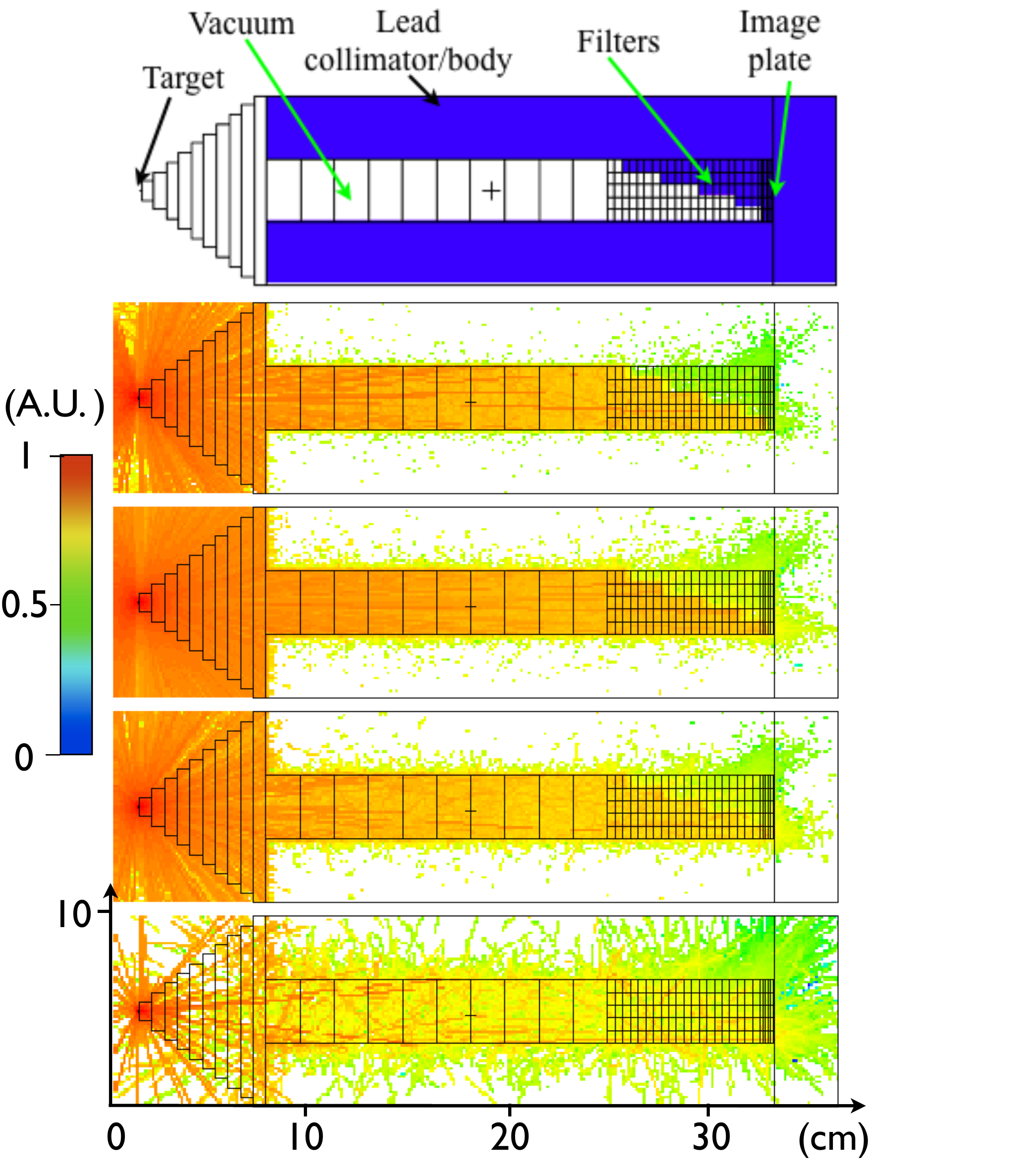}
\caption{Top: Schematic diagram of the detector. Note that the cells within the vacuum (cell walls are depicted by the black lines) are not part of the detector geometry but were used for variance reduction population control techniques. From second top to bottom the images depict relative spectrally binned photon density, the photon energy bins are: 1 to 10 keV, 10 to 100 keV, 100 to 300 keV, 300 keV to 1 MeV. For this simplified design run, the target is placed artificially close to the detector, electrons are injected into the target from the left. The bremsstrahlung photon flux shown was generated using an exponential energy spectrum of 0.5 MeV slope temperature.}
\label{fig:fluxincannon}
\end{figure}

\begin{figure}[ht]
\centering
\includegraphics[scale=0.60,trim=02mm 08mm 225mm 1mm,clip]{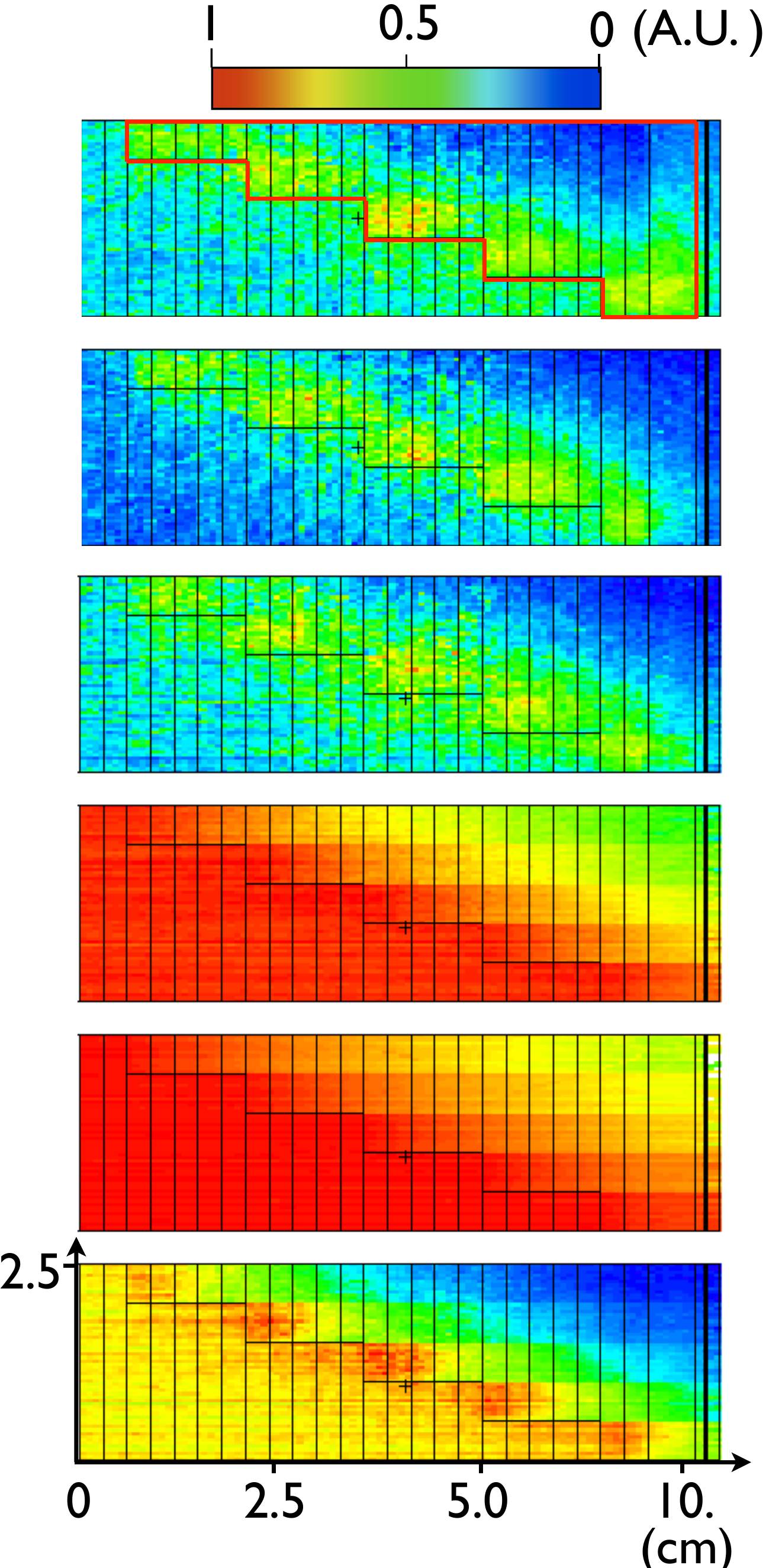}
\caption{The spatial distribution of the normalised photon density per source electron within one planar slice through the filters in various spectral bins. From top to bottom: photon flux 1 keV to 500 keV, 500 keV to 600 keV, 600 keV to 1 MeV, 1 -10 MeV, 10-100 MeV, 1 keV to 100 MeV (all the photons in the problem). In this simplified run a photon source of uniform photon energy probability up to 100 MeV with zero divergence was injected just before the entrance to the filters. Photons enter from the left into a vacuum, the Pb filters are outlined in red in the upper image. }
\label{fig:fluxinfilters}
\end{figure}

\begin{figure}[th]
\centering
\includegraphics[scale=0.42,trim=04mm 10mm 150mm 175mm,clip]{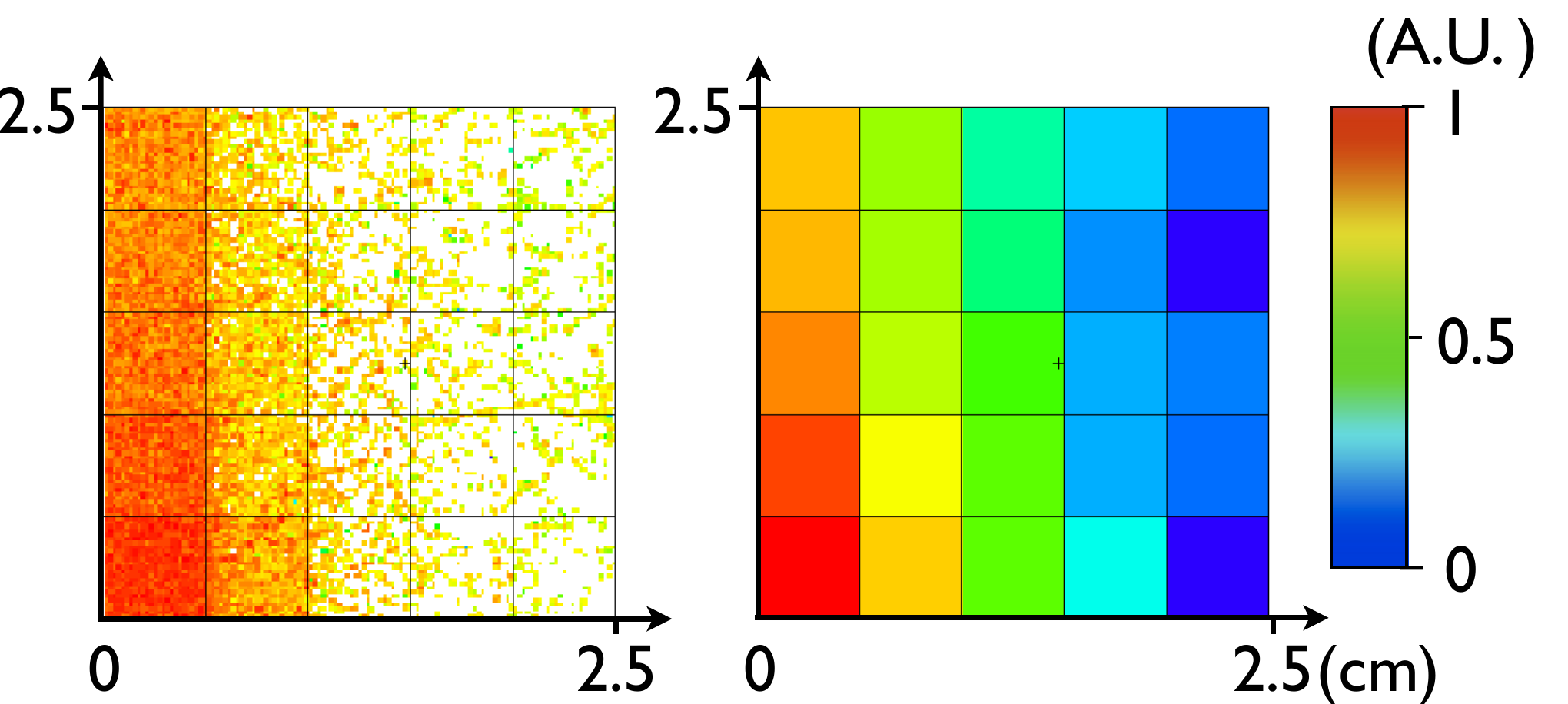}
\caption{Energy deposition within the phosphor layer of the image plate by all photons, as modelled with \sc mcnpx\rm. The filter thickness increases monotonically going from bottom to top in each column, the filters in the columns to the left are thinner than those on the right. (Left) high resolution image of energy deposition, black lines depict the layout of the 25 filters. (Right) Integrating the signal over the spatial region corresponding to one filter yields a clear signal. The filter structure is visible in both the high and low resolution images.}
\label{fig:ipedep}
\end{figure}

The laser-interaction solid target, detector and image plate were modelled in 3D in \sc mcnpx\rm, using accurate dimensions, material densities and cross-sections. During the detector design, an electron beam with spectral and spatial (both angular and position) distributions representative of those expected experimentally was injected into the target, generating bremsstrahlung photons. The spectral and spatial characteristics of these bremsstrahlung photons are related to those of the electron source. The photons propagate through the vacuum and then into the detector. Within the detector they pass through many centimetres of high $Z$ filtering before depositing their energy within the $\sim 10$ \textmu m phosphor layer of the image plate. However the probability of a bremsstrahlung photon depositing energy in the phosphor layer is extremely small, meaning the problem is effectively computationally intractable in its raw form, especially given the many design iterations required. During the conceptual design, huge computational gains were made by sub-dividing the problem and using variance, or relative error, reduction techniques \cite{Pelowitz:aa}. An example of a design run is shown in figure \ref{fig:fluxincannon}, in this simplified setup the target has been moved artificially close to the detector to improve the statistics of the Monte Carlo modelling. From figure \ref{fig:fluxincannon} it can be seen that even the lowest energy photons are able to propagate through the thick Pb filtering, although this will in part be due to the creation of low energy photons by the more penetrating high energy photons. To the left of the target it is apparent that more of the lower energy photons are scattered backwards out of the target than propagate towards the detector. There is a lower probability of a higher energy photon being created due to the (assumed) exponential electron energy distribution ($P(E)\propto \mbox{exp}(-E/kT)$) where $P$ is the probability of an electron being injected with energy $E$, $k$ is Boltzmann's constant and $T$ the slope temperature. The lower population of the highest energy photons is clearly visible in the bottom image, as is their greater ability to penetrate further through the Pb walls of the detector. Figure \ref{fig:fluxinfilters} shows the flux distribution within the filters in more detail. From these images various effects are apparent; firstly the low energy photon flux \emph{increases} within the filters, this somewhat counterintuitive result is due to K$_{\alpha}$ creation and pair production by the higher energy photons, this effect is greatly exaggerated in this run as the injected photon spectrum is flat up to 100 MeV; in more realistic distributions there are far fewer higher energy photons than lower energy photons. The images also show that the higher energy photons are less susceptible to scattering and hence will deposit energy within the correct spatial region of the image plate more effectively than lower energy photons. Figure \ref{fig:ipedep} shows that while the modelling predicts lateral photon scattering to occur within the filters, by integrating over each region (corresponding to one filter) a clear signal can be obtained. In this case the fast electron source temperature is relatively low with a slope temperature of 100 keV, hence the thickest filters are not of use - the energy deposition is relatively uniform over the region of the thickest filters. For higher fast electron temperatures it was found that the ability to discriminate between the thicker filters improves. 

From figure \ref{fig:pbfiltering} it can be seen that the data for the lowest photon energies has significantly higher relative errors:- although the error bars are not explicitly plotted for clarity, this is visible in the more erratic nature of the curves towards the bottom left of the graph. All but the thinnest filters prevent all of the 100 keV photons from propagating through the filters, meaning there is {\it no} energy deposition for 100 keV photons with thicker filters. Examination of the trends of the curves and physical intuition suggest this result is incorrect. In reality some of the 100 keV photons would have propagated through the filtering depositing their energy within the phosphor, however the computational limitations on the number of particles that can be simulated means that such an improbable event may not be sampled, and even if it is (as in the case of the thinner filters), it may be sampled so few times that the relative error is large. Monte Carlo population control techniques allow the 5 dimensions of phase space (3 space, energy, and time) to be subdivided with an `importance' $n$ applied to each region. As photons propagate through the filters (and hence phase space) their population will reduce exponentially and be shifted to lower energies. Simultaneously the pre-set importances will cause the photons to be split into $n$ photons of weight $1/n$. In this way a large number of low weight, low energy photons can be tracked through the Pb, ultimately depositing some energy in the image plate. Even though the energy deposited per tracked particle is low, the statistics are good as many particles deposit an appropriate amount of energy. To make the problem tractable, a new photon population control technique was devised which allowed the user a far greater degree of control over the spatial distribution of the importances than is normally possible with \sc mcnpx\rm. This involved subdivision of the filter geometry into smaller cells, the `importances' of the subcells reflect the photon propagation through each of the filters, hence the exponential reduction in photon number as a function of detector thickness is simultaneously counteracted by splitting these photons, creating more photons of low `weight'. In combination with automated input deck writing routines, this enabled a highly optimized solution to this specific transport problem to be implemented that was otherwise essentially intractable using pre-exisiting variance reduction techniques. 

The detector Pb wall thickness was designed such that the K$_{\alpha}$ radiation produced within the walls of the experimental chamber would be reduced by $\sim10^{-10}$. The collimator length was minimized in order to save space. Its length was based on \sc mcnpx \rm modelling which showed that even 100 MeV photons are extremely unlikely to propagate through 20 cm of Pb. 

The three dimensional filtering array is potentially susceptible to cross-talk between the various filters due to lateral Compton scattering within the filters as the photons propagate towards the image plate detector, excessive scattering could have rendered the design unusable. This cross-talk was reduced geometrically by minimizing the required filter depth by maximizing the total photon cross-section of the filter material (within practical limitations) by using Pb filtering. The modelled energy deposition in the image plate is shown in figure \ref{fig:ipedep}, it can be seen that there is a reasonably clear distinction between the signals formed on each region corresponding to a channel of the detector.

A 0.4 T magnet was positioned between the laser target and the bremsstrahlung detector to bend those electrons which do escape the target away from the mouth of the detector. This field deflects 100 MeV electrons away from the detector entrance if the magnet is located 40 cm in front of the detector.  

\section{Fast Electron Temperature, Absorption and Divergence Modelling}
\label{sec:bremmodel3}

This description refers to the modelling performed for the experiment described in section \ref{expt} as an example. It is typical of that required for the analysis of laser-solid interactions using this diagnostic. The experimental setup was modelled in 3D within \sc mcnpx \rm including all of the target layer materials, $n$ bremsstrahlung detectors are positioned exactly as per the experiment both in terms of angle and distance from the source. A probabilistic distribution of the source electron energies was assummed, corresponding to a 3D single-temperature relativistic Maxwell-Jutner distribution of the form: 
\begin{equation}
f(\gamma,\theta)\propto f(\theta)\gamma^2\beta \frac{m_ec^2}{kT}\mbox{exp}(-\gamma\frac{m_ec^2}{kT})
\end{equation}
where $f(\theta)$ is an arbitrary assumed fast electron angular distribution function defined relative to the injection axis, $\gamma=1/\sqrt{1-\beta^2}$, $\beta = v/c$, $c$ is the speed of light, and $m_e$ and $v$ are the electron mass and speed respectively. In order to make the problem computationally tractable, it is subdivided, the modelling steps are as follows: 

\begin{figure}[htb!]
\centering
\subfigure[]{}\includegraphics[scale=0.25,angle=0,trim=0mm 0mm 0mm 0mm,clip]{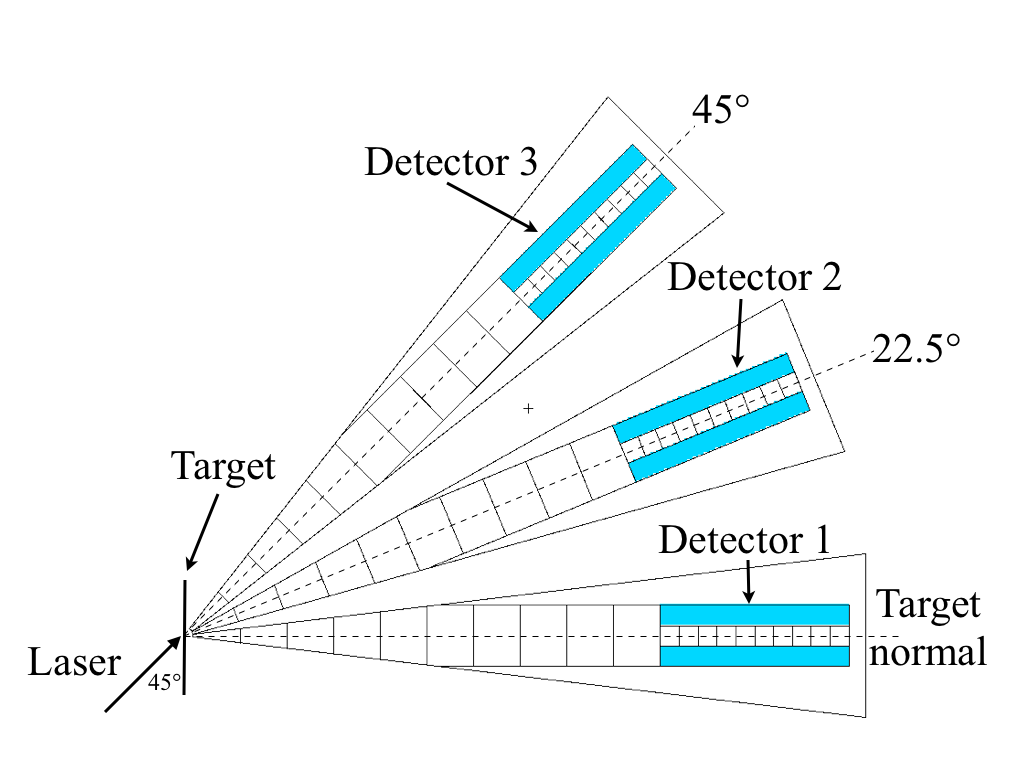}
\caption{The 3D \sc mcnpx \rm model with 3 detectors positioned as per the experiment described in section \ref{expt}.}
\label{fig:mcnpxabssetup}
\end{figure}

\begin{enumerate}
\item
Using a single detector a range of Maxwellian fast electron sources with temperatures varying from 0.1 MeV to 0.8 MeV were `injected' into the target in \sc mcnpx \rm (the temperature range was based on the experimental intensity and using Beg's\cite{Beg:1997pd} and Wilks'\cite{Wilks:1997aa} scalings). The modelled photon energy deposition within the phosphor layer of the image plate was compared to the experimentally measured PSL value using absolute calibration values from the image plates \cite{Scott:2011fk}. The modelled result which had the best least squares fit to the experimental data was deemed to be the correct relativistic Maxwellian temperature.
\item
Using the fast electron spectrum found to best fit the data (step 1) and an assumed fast electron angular divergence, a run is performed using $n$ detectors each of which is positioned at different angles as per the experiment (see fig. \ref{fig:mcnpxabssetup}). The spectrum of bremsstrahlung photons entering each detector is recorded. 
\item
$n$ new input decks are created, one for each detector. Each deck uses the bremsstrahlung spectrum (from step 2) to define the spectrum of the photons injected in this run. This calculates the coupled photon-electron transport through $n$ detectors, their filters and ultimately the energy deposited within the phosphor layer of the image plate per bremsstrahlung photon entering the detector. 
\item
Using the absolute calibration of the image plates the modelled energy deposition within the phosphor layer of the image plate is compared with the experimentally measured values. Depending on the result of this comparison either the electron source spectrum and/or divergence may be iterated until a good match is found. 
\item 
The total energy in the fast electron distribution is calculated as follows: 

(a) The modelled bremsstrahlung spectrum entering each of the detectors is integrated over each detector cavity and energy spectrum, giving the total number of photons entering each detector \emph{per source electron} entering the target. 

(b) The total modelled energy deposited in the image plate phosphor layer \emph{per source photon} is found by integrating over the phosphor volume. Using (a) this is then converted to total energy deposited in the phosphor layer \emph{per source electron}. 

(c) Using the image plate absolute calibration, the total experimentally measured energy deposition in units of PSL are converted to units of MeV. 

(d) From (b) and (c) the number of source electrons required to create the experimentally measured signal are calculated. 

(e) The mean electron energy of the fast electron distribution is calculated given its temperature and distribution. The total energy of the fast electron distribution is found by multiplying the number of electrons from (d) by the mean energy of the fast electron distribution.

\end{enumerate}
\begin{figure}[t!!!]
\centering
\includegraphics[scale=0.4,angle=0,trim=30mm 10mm 30mm 4mm,clip]{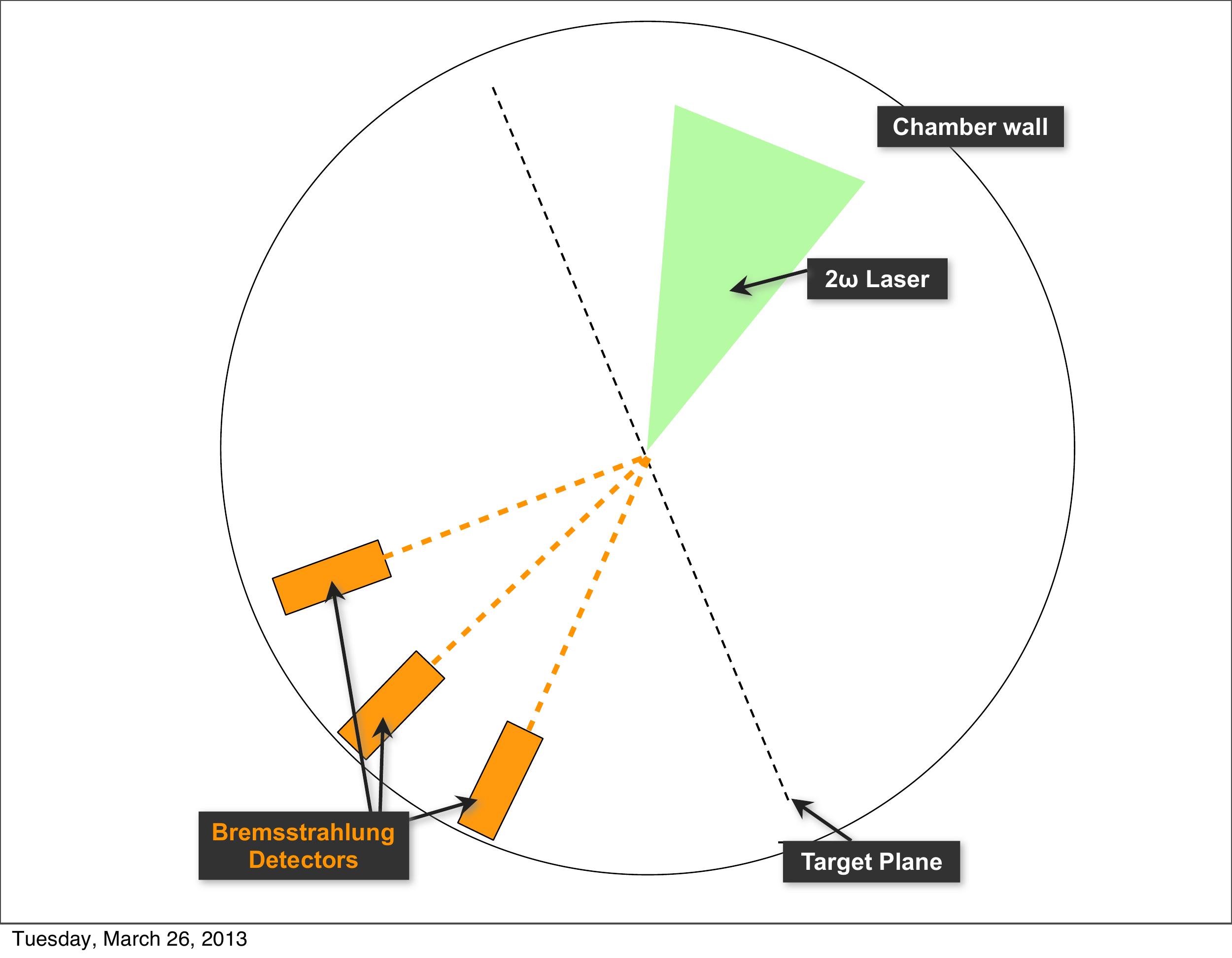}
\caption{The experimental setup. }
\label{fig:2w_expt_setup}
\end{figure}

When calculating the laser energy absorbed into fast electrons it is crucial that the modelled response of all the detectors matches that of the experimental data, as a consequence some fast electron divergence (or potentially angularly varying fast electron temperature) of the electron source must be assumed. Conversely, the value of fast electron divergence which best fits the data is a measure of the experimental fast electron divergence. This measurement of divergence contains spectral information but is very limited in angular resolution compared to existing techniques due to the limited number of detectors. Sensitivity to the assumed divergence angle will only occur when the fast electron temperature is sufficiently high such that the bremsstrahlung is confined to the 1/$\gamma$ angular cone. Therefore for lower fast electron temperatures the inferred fast electron temperature is insensitive to the assumed fast electron angular distribution function (e.g. for $T=140$ keV the 1/$\gamma$ half angle is $\sim$ 45$^{\circ}$). Similarly lower fast electron temperatures also mean no effective measure of the fast electron divergence can be made using this technique. Bremsstrahlung theory leads us to expect anisotropic distributions from higher fast electron temperatures and hence should occur during experiments with higher values of $I\lambda^2$. It should also be noted that because the target producing the bremsstrahlung radiation only covers the region of solid angle in the forward direction with respect to the incident laser, this technique provides only measures laser energy absorbed into fast electrons which are forward-going. 

\section{Frequency Doubled, Relativistically-Intense, Laser-Solid Experiment}
\label{expt}

\begin{figure}[htb!]
\centering
\subfigure[]{}\includegraphics[scale=0.52,angle=0,trim=15mm 60mm 32mm 80mm,clip]{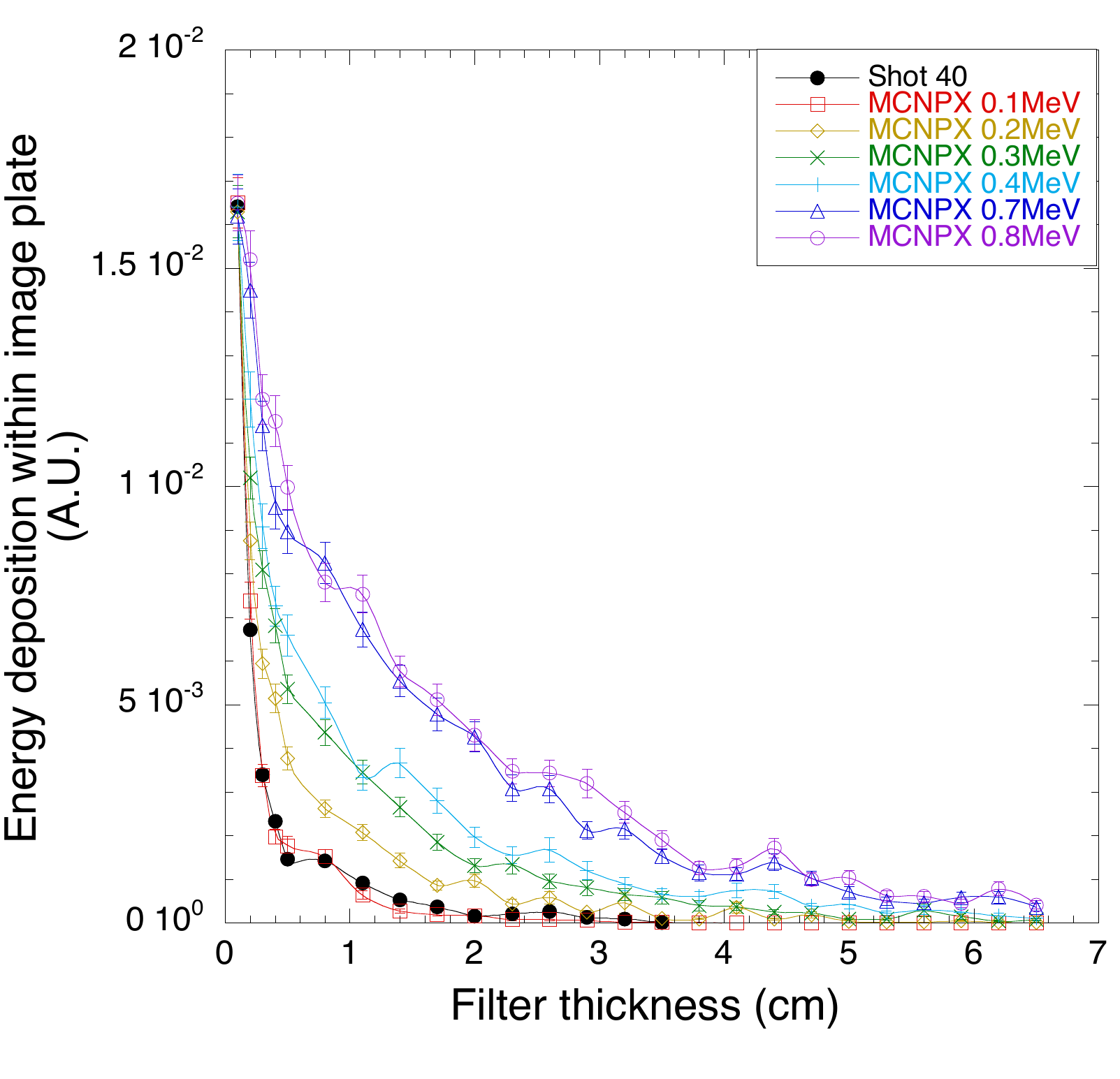}
\subfigure[]{}\includegraphics[scale=0.52,angle=0,trim=15mm 60mm 32mm 80mm,clip]{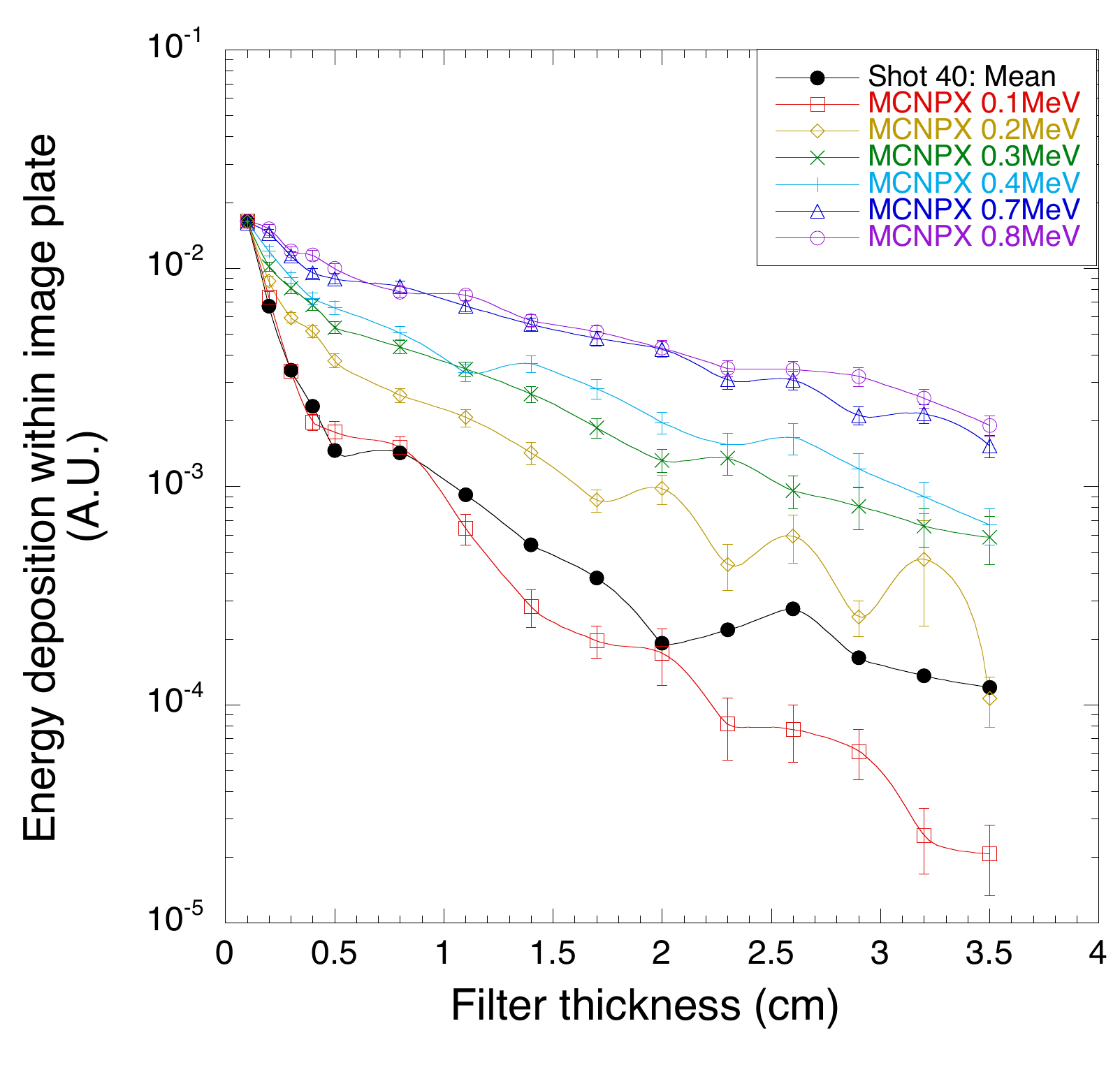}
\caption{(a) \& (b) show the same data plotted on linear and log scales. The data is compared to the modelled detector response for various initial relativistic Maxwellian electron temperatures. The experimental data is the mean PSL value within each `region' of the image plate, a region exactly corresponds to the shadow cast by one of the filters onto the image plate, likewise the \sc mcnpx \rm value is the mean energy deposition within the same location and area of the modelled image plate's phosphor layer. The lines shown are simply meant to guide the reader's eye.}
\label{fig:shot40approxtemp}
\end{figure}

An experiment was performed at the LULI 2000 laser facility at the Ecole Polytechnique in Paris in order to characterize the properties of the fast electron beam generated when relativistically-intense frequency-doubled laser light interacts with a solid target. On this experiment 25J of 527 nm light was incident on the target at 45 degree p-polarization, with a pulse length of 800 fs and spatial full-width-at-half-maximum of 15 \textmu m yielding a peak intensity of $9\times 10^{18}$ W/cm$^2$ and mean intensity (found by spatially and temporally averaging the Gaussian distributions over the spatial and temporal FWHM repectively) of $5\times 10^{18}$ W/cm$^2$. For the purposes of absolute laser absorption measurements the targets were composed of successive layers Al (10 \textmu m), Ag (10 \textmu m), Au (10 \textmu m), Cu (10 \textmu m), Al (10 \textmu m) and CH (300 \textmu m), with the laser incident on the first Al layer. Absolute bremsstrahlung radiation measurements require the prevention of refluxing as multiple transits of the electrons within the target will cause multiple bremsstrahlung emissions from within the target. By attaching a 300 \textmu m mylar layer to the target rear, the majority of the fast electrons are stopped within 600 \textmu m of plastic (based on Monte Carlo modelling of an appropriate fast electron energy distribution), meaning the experimental signal should be a good approximation to a single pass of the fast electrons passing through the high $Z$ material.

\begin{figure}[htb!]
\centering
\subfigure[]{}\includegraphics[scale=0.35,angle=0,trim=115mm 47mm 75mm 80mm,clip]{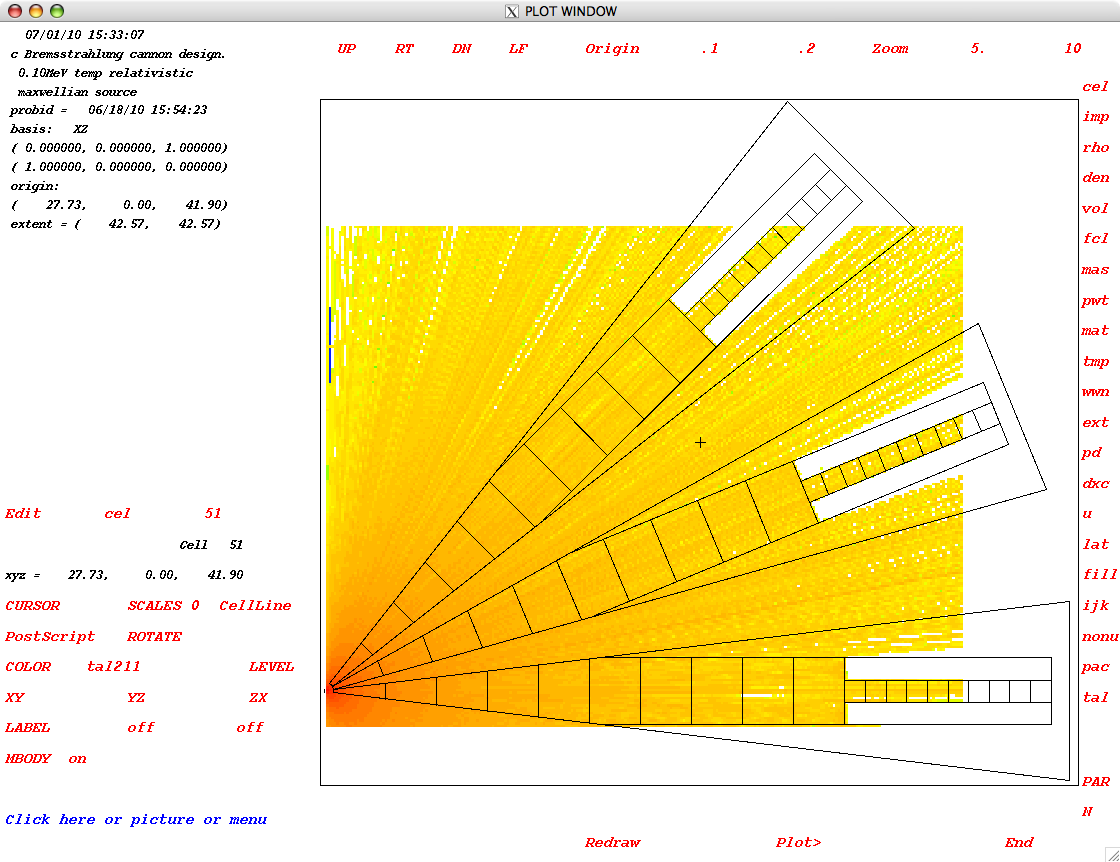}
\subfigure[]{}\includegraphics[scale=0.45,angle=0,trim=10mm 60mm 0mm 80mm,clip]{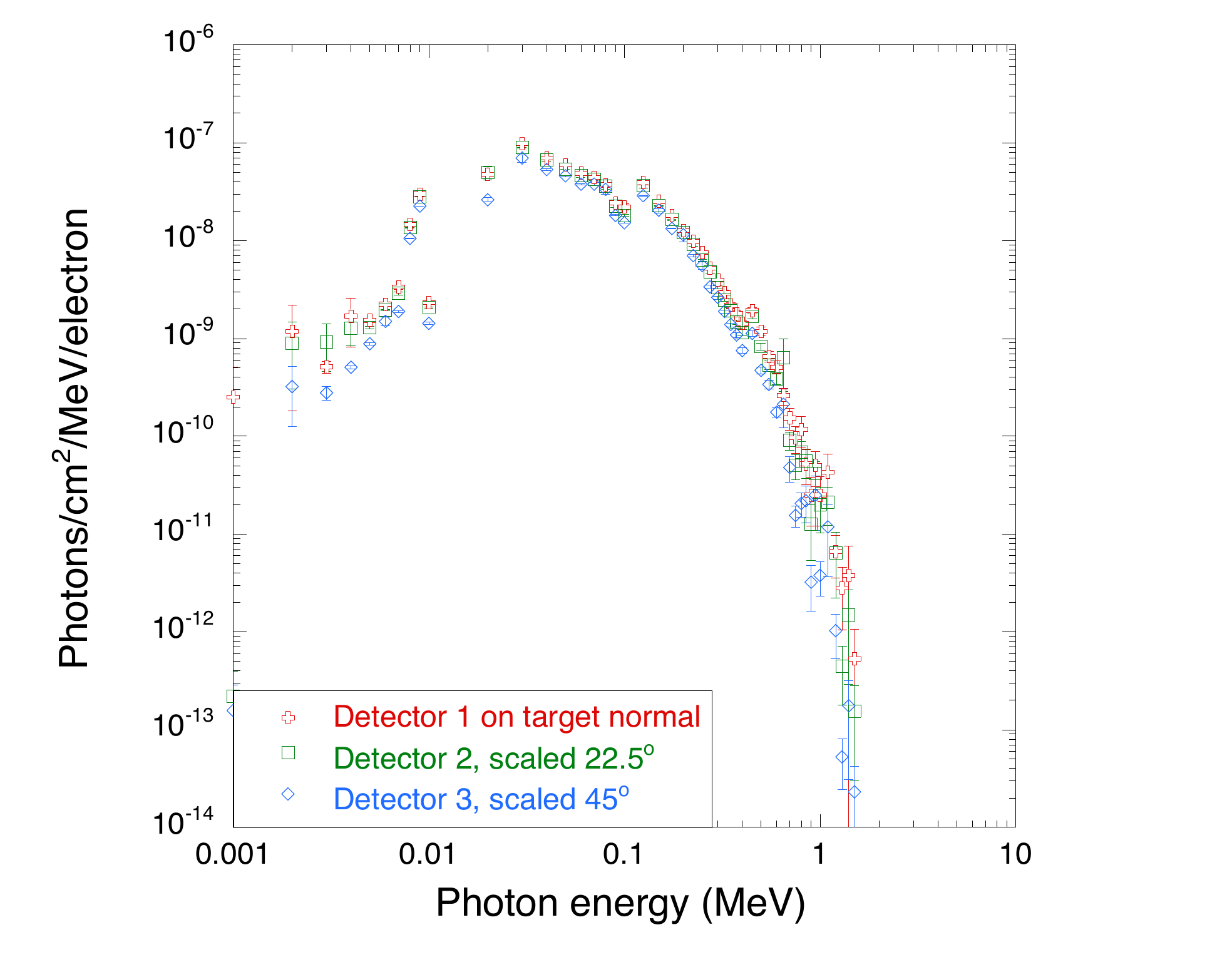}
\caption{(a) The modelled angular distribution of bremsstrahlung flux of all energies. The target is at the bottom left of the image, while the lead shielding of the three detectors is visible as the white regions (low photon flux)towards the top and right. (b) The modelled flux spectra measured at the entrance to the three bremsstrahlung detectors (scaled to account for the differences in the detector distances from the target) shows very little spectral variation with angle.}
\label{fig:fluxangledist}
\end{figure}

The bremsstrahlung spectra and angular distribution were measured using three of the detectors described in section \ref{sec:canondesign} as shown in figure \ref{fig:2w_expt_setup}. Based on the work of Santala \emph{et al} \cite{Santala:2000lp} it was anticipated that the peak emission should be located along the target normal direction given the expected high experimental contrast, hence they were positioned along the laser axis and then at two 22.5$^{\circ}$ increments with the third detector at target normal. Magnets were positioned such that all electrons below 40 MeV were prevented from entering the detectors.

\subsection{Fast Electron Energy Spectrum} 

The procedure to find the fast electron energy spectrum (as outlined in section \ref{sec:bremmodel3}) was followed using the experimental target materials, densities, geometry and detector geometry modelled in 3D within \sc mcnpx\rm. Figure \ref{fig:shot40approxtemp} (a) and (b) show the relative energy deposition within the different spatial regions of the image plate (each region corresponds to an energy bin) as a function of fast electron temperature. The experimentally measured values are also over-plotted. The best fit to the experimental data was a fast electron temperature of $125\pm25$ keV. For the purpose of evaluating the fast electron temperature only the relative values of the energy deposition in each bin are important, hence arbitrary units are used. 

\subsection{Fast Electron Divergence}

The bremsstrahlung emission measured over the 45$^\circ$ measurement angle was essentially uniform. The bremsstralung emission cone angle is approximately $1/\gamma$, for 100 keV electrons this cone angle will be of the same order as the angle covered by the detectors, meaning no information about the electron angular distribution can be inferred from these experimental measurements, with this low fast electron temperature. This was confirmed by modelling the target and all three detectors in \sc mcnpx \rm using the fast electron temperature derived from the experiment. The \sc mcnpx \rm model geometry is shown in figure \ref{fig:mcnpxabssetup}, while figure \ref{fig:fluxangledist} shows the calculated bremsstralung emission over the forward hemisphere, which is essentially uniform. This modelling was performed with a $\it collimated$ electron beam entering the target with a slope temperature of 125 keV, confirming that for this relatively low fast electron temperature, the fast electron temperature is insufficient for the emitted bremsstrahlung to be measurably beamed in the direction of the propagating electrons.

\section{Summary}

In summary, a novel bremsstrahlung photon detector has been designed and developed using the 3D Monte Carlo code \sc mcnpx\rm. It has been successfully fielded on an experiment using relativistically-intense frequency-doubled laser light. Using 3D Monte Carlo modeling, the fast electron temperature was backed-out from the experimental data, it was found to be best fit by a relativistic Maxwellian of temperature $125\pm25$ keV. 

%\bibliography{Robbies_bibtex_References}

\end{document}